\documentclass[doublecol,linenumbers]{epl2}
\usepackage{amsmath}
\title{Influence of defects on the effective electrical conductivity of a monolayer produced by random sequential adsorption of linear $k$-mers onto a square lattice}
\shorttitle{Influence of defects on the electrical conductivity of a monolayer}
\author{Yu. Yu. Tarasevich\inst{1} \and  V. V. Laptev\inst{2,1} \and  V. A. Goltseva\inst{1} \and  N. I. Lebovka\inst{3,4}}
\shortauthor{Yu. Yu. Tarasevich \etal}
\institute{
\inst{1}Astrakhan State University, 20A Tatishchev Street, Astrakhan, 414056, Russia\\
\inst{2}Astrakhan State Technical University, 16 Tatishchev Street, Astrakhan, 414025, Russia\\
\inst{3}F. D. Ovcharenko Institute of Biocolloidal Chemistry, NAS of Ukraine, 42 Boulevard Vernadskogo, 03142 Kiev, Ukraine\\
\inst{4}Taras Shevchenko Kiev National University, Department of Physics, 64/13 Volodymyrska Street, 01601 Kyiv, Ukraine\\
}
\pacs{72.80.Ng}{Disordered solids}
\pacs{05.10.Ln}{Monte Carlo methods}
\pacs{68.35.Rh}{Phase transitions and critical phenomena}

\abstract{
The effect of defects on the behaviour of electrical conductivity, $\sigma$, in a monolayer produced by the random sequential adsorption of linear $k$-mers (particles occupying $k$ adjacent sites) onto a square lattice is studied by means of a Monte Carlo simulation. The $k$-mers are deposited on the substrate until a jamming state is reached, i.e. a state where no one additional particle can be placed because the presented voids are too small or of inappropriate shapes. The presence of defects in the  lattice (impurities) and of defects in the  $k$-mers with concentrations of $d_l$ and $d_k$, respectively, is assumed. The defects in the lattice are distributed randomly before deposition and these lattice sites are forbidden for the deposition of $k$-mers. The defects of the $k$-mers are distributed randomly on the  deposited $k$-mers. The sites filled with $k$-mers have high  electrical conductivity, $\sigma_k$,  whereas the empty sites, and the sites filled by  either  types of defect have a low electrical conductivity, $\sigma_l$, i.e., a high-contrast, $\sigma_k/\sigma_l \gg 1$, is assumed. We examined isotropic (both the  possible $x$ and $y$ orientations of a particle are equiprobable) and anisotropic (all particles are aligned along one given direction, $y$) deposition. To calculate the  effective electrical conductivity, the monolayer was presented as a random resistor network (RRN) and the Frank--Lobb algorithm was used. The effects of the concentrations of defects $d_l$ and $d_k$ on the electrical conductivity for the values of $k = 2^n$, where $n =1,2,\dots, 5$, were studied. Increase of both the $d_l$ and $d_k$ parameters values resulted in decreases in the value of $\sigma$  and the suppression of percolation. Moreover, for anisotropic deposition the electrical conductivity along the $y$ direction was noticeably larger than in the perpendicular direction, $x$. Phase diagrams  in the ($d_l, d_k$)-plane for different values of  $k$ were obtained.
}

\begin{document}

\maketitle

\section{Introduction: electrical conductivity of inhomogeneous media}
The physical properties of inhomogeneous media have attracted significant attention  in the scientific community since the 19th century~\cite{Maxwell1881}. Mainly, those  efforts have been  concentrated on  binary inhomogeneous materials. One of the main problems in the theory of disordered systems is the calculation of the electrical conductivity for a random mixture of insulating and conducting materials~\cite{Clerc1990AdPhys}. In particular, the singular behaviour of the electrical conductivity near a percolation threshold is of interest~\cite{Efros1976pssb}. Investigations of the physical properties of inhomogeneous media are significant for  numerous applications such as the production and use  of nanocomposites~\cite{McLachlan2007JNM}. Theoretical prediction of the effective properties for multiphase material systems is very important for the  analyses  of material performance and for the design of  new materials~\cite{Wang2008MSE}.

An inhomogeneous medium can be considered as either  continuous or discrete. Accordingly, two complementary approaches are used to describe the electrical properties of such  disordered media, i.e. the continuous approach and the discrete approach. The continuous  approach originates from Maxwell's works. In a Maxwell approximation,
the impurities are supposed to be at a low concentration and to have  a regular compact form, e.g. sphere or ellipsoid (``\dots spheres \dots placed in a medium \dots at such distances from each other that their effects in disturbing the course of the current may be taken as independent of each other''~\cite[p.~440--441]{Maxwell1881}, thus the mixture is diluted. The Maxwell approximation implies a linear dependence of the electrical conductivity on the concentration of inclusions. An extended  approximation obtained in terms of the Maxwell approach allows the electrical properties of composites to be described for a wide concentration range and even demonstrates  the presence of the percolation threshold~\cite{Snarskii2007PhysU}.

One widely used approach is the effective medium approximation (EMA)~\cite{Bruggeman1935AnnPhys}.  The classical EMA provides  a good description of the  physical properties for any concentration except for a narrow range  around the percolation threshold~\cite{Snarskii2007PhysU}. At present, a  more advanced version of the EMA, known as the generalized effective medium approximation (GEMA)~\cite{McLachlan2007JNM}, offers  fairly good description of the physical properties even near the percolation threshold. An alternative description, i.e. the  percolation approach, has been applied to a system consisting of randomly distributed metallic and dielectric regions~\cite{Efros1976pssb}. Notice, in the percolation approach, an inhomogeneous material  can be treated both as a continuous  and as a discrete medium. The general percolation problem of cutting randomly centred insulating holes of arbitrary shape in a two-dimensional conducting sheet and its electrical conductivity has been investigated~\cite{Garboczi1991PRA}. The review~\cite{Clerc1990AdPhys} is devoted to the AC electrical response of binary inhomogeneous materials, modelled as bond percolation networks.
Percolation and the EMA, as they apply to the electrical conductivity of composites, are reviewed in~\cite{McLachlan1990JACE}.

A special group  of inhomogeneous media are flat (2D) systems. Simulation of the electrical properties of 2D inhomogeneous systems (thin films) is motivated by their numerous   applications. The resistance of a two-dimensional system of conducting sticks depends on systems anisotropy~\cite{Balberg1983SSC}.
It was shown that the conductivity of a two-phase thin film, with both  equal concentrations of the phases and their random distribution, is equal to the geometric mean of the conductivity of the phases~\cite{Dykhne1971JETP}. The effective  conductivity of random two-phase flat systems has been  studied using an approach that differ from the effective medium approximation~\cite{Bulgadaev2003PLA}.

The physical  properties of monolayers produced by RSA have been  widely studied and  discussed with special attention being paid to the  effects of defects~\cite{Kondrat2006JChPh,BudinskiPetkovic2016JSM,Centres2015JSMTE} and particle size distribution~\cite{Hart2016PRE,Kuriata2016MTS}. Recently, the percolation behaviour of the effective conductivity for a  lattice model with interacting particles was reported~\cite{Wisniowski2016PhysA}. Percolation and jamming phenomena have been  investigated for the random sequential adsorption (RSA) of dimers on a square lattice, where  the influence of dimer alignment on the electrical conductivity was examined~\cite{Cherkasova2010EPJB}. A systematic study of the  electrical conductivity of a monolayer produced by the RSA of linear $k$-mers (with values of $k$ up to 128) onto a square lattice was, additionally, performed by means of computer simulation~\cite{Tarasevich2016PRE}.

In real-world systems, the surfaces may be chemically heterogeneous and contain defects~\cite{Adamson1997}, or may be prepatterned~\cite{Cadilhe2007JPhysCM}. The structure of  elongated particles, e.g., carbon nanotubes, may also be highly heterogeneous and contain insulating regions, e.g. due to irradiation damage~\cite{Ritter2010} or chemical functionalization~\cite{Wepasnick2010}. The percolation and jamming of $k$-mers on disordered (or heterogeneous) substrates with defects, or of $k$-mers with defects, have also attracted a great deal of attention~\cite{Ben-Naim1994JPhysA,Lee1996JPhysA,Kondrat2005,Kondrat2006JChPh,Cornette2003epjb,Cornette2006PLA,Cornette2011PhysA,Budinski-Petkovic2011,Budinski-Petkovic2012,Tarasevich2015PRE, Lebovka2015PRE}.

A generalized variant of the RSA model where both the  $k$-mers and the  lattice have  defects has been proposed~\cite{Lebovka2015PRE}. Some of the occupying $k$ adjacent sites are considered as insulating and some of the lattice sites are occupied by defects (impurities). In this model, even a small concentration of defects can inhibit percolation for relatively long $k$-mers. Recently, some results concerning percolation and the  electrical conductivity of monolayers produced by the RSA of aligned linear $k$-mers with defects onto a square lattice with impurities have been  presented~\cite{Tarasevich2016JPhCS}.

In this paper, we quantitatively examine the electrical conductivity of monolayers, paying special attention to the influence of defects on the electrical properties.  We consider the monolayers  as random resistor networks (RRN).

\section{Methods}\label{sec:methods}
In our computer simulation, we utilized RSA~\cite{Evans1993RMP} to produce a monolayer. We employed a discrete two-dimensional substrate, namely a square lattice with periodic boundary conditions (a torus). The linear $k$-mers, i.e. particles occupying $k$ adjacent lattice sites, were randomly deposited on the substrate. The values of $k$ were $2^n$, where $n=1,2,\dots,5$. Some fraction of the lattice sites ($d_l$) may be forbidden for the deposition of objects. We treated these sites as defects or impurities. These impurities had no effect on the electrical conductivity of the substrate but did affect the  deposition of particles. The linear $k$-mers were randomly deposited on the substrate until a jamming state occurred, i.e. a  state when no one additional particle can be placed because the presented voids are too small or of inappropriate shape. We examined the isotropic as well anisotropic deposition of the particles. During isotropic deposition, both the possible orientations, $x$ and $y$, of a particle are  equiprobable. During anisotropic deposition, all particles are aligned along one given direction, $y$. Overlapping with previously deposited particles was strictly forbidden, as a result, a monolayer was formed. Adhesion between the particles and the substrate was assumed to be very strong, so once deposited, a  particle could not slip over the substrate or leave it (diffusion and detachment of the  particles were impossible).

As a  first step,  point defects (impurities) were randomly embedded in the  lattice sites up to a  given concentration $d_l$. After that, $k$-mers with identical electrical properties of all their sites were deposited onto the substrate using the RSA algorithm until the jamming state was reached. Finally, defects were added to these deposited particles, i.e., some randomly chosen $k$-mer sites were marked as insulating. We studied how the effective electrical conductivity varied  with the concentration of defects, $d_k$.

\begin{figure}[htbp]
  \centering
  \includegraphics[width=0.75\linewidth]{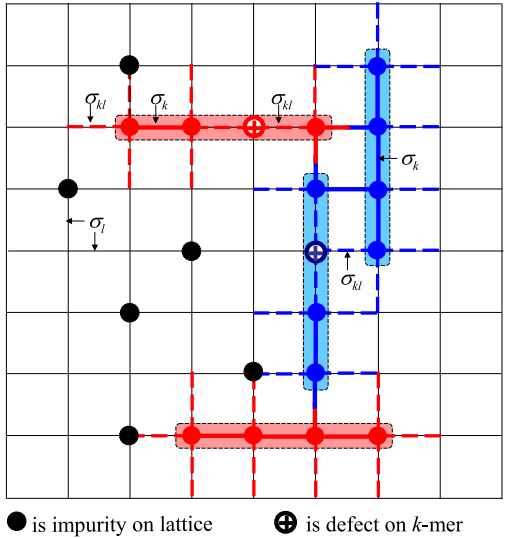}
  \caption{Fragment of a square lattice with four deposited 4-mers (horizontal and vertical). Impurities on the lattice are shown by black circles, and defects on k-mers are indicated by crosses. Different electrical conductivities of the bonds between empty sites, $\sigma_l =1$  (thin lines),  filled sites,
$\sigma_k =10^6$ (thick lines), and empty and filled sites,
$\sigma_{kl} =2\sigma_k\sigma_l/(\sigma_k+\sigma_l)\approx 2\sigma_l$ (dashed lines) were assumed.
}\label{fig:lattice}
\end{figure}

Figure~\ref{fig:lattice} presents a  fragment of a square lattice with four deposited 4-mers (horizontal and vertical). Impurities on the lattice are shown by black circles, and defects on $k$-mers are indicated by crosses. Different electrical conductivities of the bonds between the empty sites, $\sigma_l =1$  (thin lines),  filled sites, $\sigma_k =10^6$ (thick lines), and  empty and filled sites,
$\sigma_{kl} =2\sigma_k\sigma_l/(\sigma_k+\sigma_l)\approx 2\sigma_l$ (dashed lines) were assumed.

To find the effective electrical conductivity, the torus was unrolled into  a plane and two conducting buses were applied to its opposite sides. The electrical  conductivity of the resulting RRN was calculated between these buses using the Frank-Lobb algorithm~\cite{Frank1988PRB}. This RRN is an image of the original monolayer, it has a regular structure but randomly distributed conductivities. A  preliminary scaling analysis of the electrical conductivity behaviour at different values of $k$ and $L$ has recently been performed for the defect-free problem~\cite{Tarasevich2016PRE}.  The difference between the approximated value of electrical conductivity in the limit of the infinite system and $L=100k$ was of the order of several percent. This is the  reason why in our computations, for any value of $k$, the lattice size $L$ was $L = 100k$. For each given value of $k$, the computer experiments were repeated 10 times, then, the logarithm of the effective electrical conductivity was averaged.

For each value of $k$, we studied three situations. We examined one limiting case when defects were completely absent from the  substrate  ($d_l=0$). We also considered an  opposite limiting case, corresponding  to a situation where the  defect concentration on the  lattice was so  high that the deposited particles could not form a chain wrapping the substrate, hence, the monolayer was  insulating even at the jamming concentration. In other words, jamming coverage was  of the order of the percolation threshold. We also took values of $d_l$ slightly  smaller than  this value.
Additionally, we studied systems with some intermediate concentrations of defects on the lattice.

\section{Results}\label{sec:results}
\begin{figure}[htbp]
  \centering
  \includegraphics[width=\linewidth]{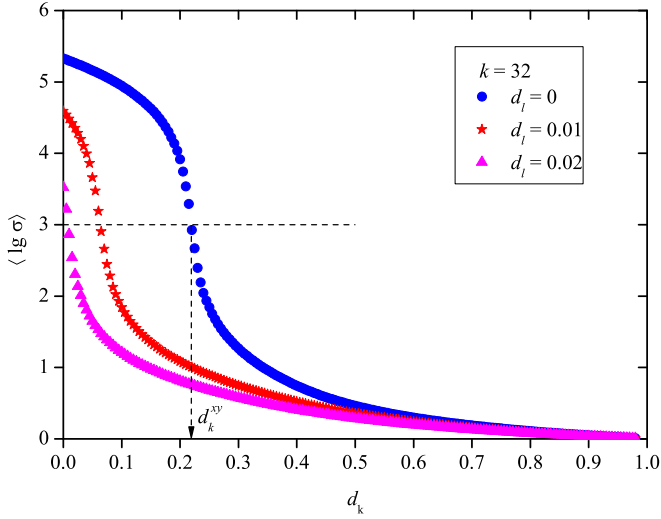}
  \caption{Electrical conductivity, $\sigma$, vs  defect concentration on the $k$-mers, $d_k$, for different values of defect concentration, $d_l$, on the lattice. The arrow  shows an example of the critical ``geometrical'' concentration $d_k^{xy}$ for $d_l=0$. The isotropic deposition, $k=32$, $L=100k$, results are averaged over 10 independent runs. The statistical error is of the order of the marker size.}\label{fig:Conductivity32i}
\end{figure}

Upon the isotropic deposition of particles, the effective electrical conductivity of a monolayer decreased  as the concentration of insulating defects on the particles increased.
Figure~\ref{fig:Conductivity32i} presents an example of such dependencies $\sigma(d_k)$ for different values of $d_l$ and for $k=32$. For other value of $k$, the behaviour of the electrical conductivity was  very similar to the behaviour presented in Figure~\ref{fig:Conductivity32i}. When any lattice site is allowed for deposition ($d_l=0$), the curve $\sigma(d_k)$ is a typical sigmoid.

At the given value of $d_l$, the observed transition of electrical conductivity $\sigma(d_k)$ from the high-conducting to non-conducting state was fairly smooth  and corresponded to the  behaviour of the order parameter in a  second-order phase transition in  the  presence of an external field. In the problem under consideration, the reciprocal electrical contrast $h=\sigma_l/\sigma_k$ plays the role of the ``external field'', $h \ll 1$. The external field smears the phase transition~\cite{Snarskii2016}. An infinitely large electrical contrast, when  $h =0$, corresponds to the absence of an  external field.

For isotropic deposition, the critical ``geometrical'' concentrations $d_k^{xy}$ that correspond to the points of mean geometric conductivity
\begin{equation}\label{eq:geom}
 \sigma_g = \sqrt{\sigma_k \sigma_l}
\end{equation}
are fairly close to the percolation thresholds.  This  corresponds exactly to  the  prediction for 2D systems in the case of systems with equal concentrations of the phases\cite{Dykhne1971JETP}.

When the concentration of impurities on the lattice is so  large that it almost blocks the formation of a spanning cluster ($d_l=0.02$), the effective electrical conductivity drops from $\sim10^3$ to $\sim1$ without a visible inflection point. The behaviour of the effective electrical conductivity confirms that, in the case of the deposition of long particles, even a very small concentration of defects on the substrate can prevent the formation of a conducting chain of particles.

\begin{figure}[htbp]
  \centering
  \includegraphics[width=\linewidth]{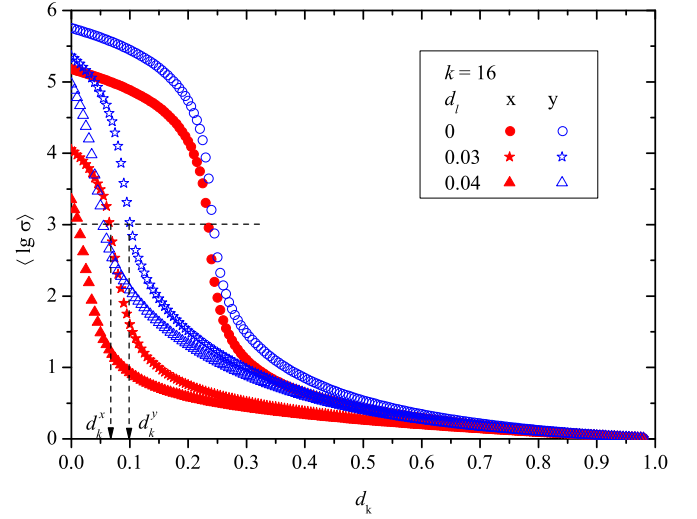}\\
  \includegraphics[width=\linewidth]{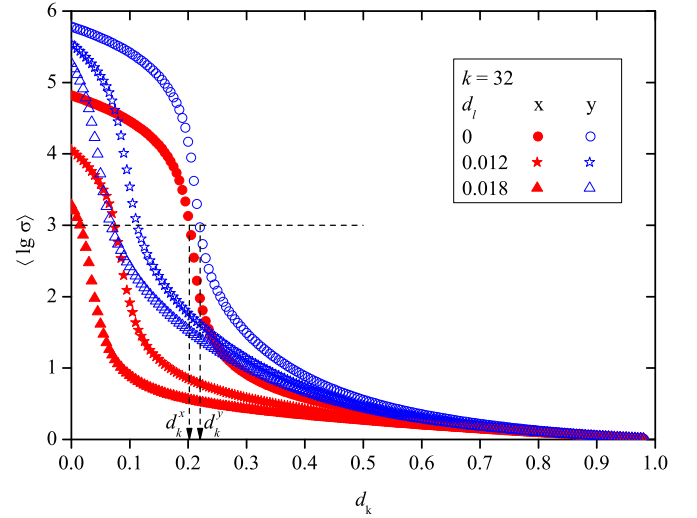}
  \caption{Electrical conductivity, $\sigma$, vs  defect concentration on the $k$-mers, $d_k$, for different values of defect concentration on the lattice, $d_l$. The anisotropic deposition, a) $k=16$, b) $k=32$, $L=100k$,  results are averaged over 10 independent runs.  The statistical error is of the order of the marker size. The arrows  show the examples of the critical ``geometrical'' concentration $d_k^x$ and $d_k^y$ in the $x$ and $y$ directions.}
   \label{fig:Conductivity8-32a}
\end{figure}

With anisotropic deposition of the particles onto the lattice, the  behaviour of the effective electrical conductivity is much more surprising. In this case, all the deposited particles are aligned along the $y$ axis.  As expected, the longitudinal effective electrical conductivity (i.e., the conductivity, measured along the $y$ axis, $\sigma_y$) and the transversal one (i.e., the conductivity, measured along the $x$ axis, $\sigma_x$) may differ. This effect has  recently been reported for the case when all  kinds of defects are  absent~\cite{Tarasevich2016PRE}. It was more pronounced for long particles (for large values of $k$). In Figure~\ref{fig:Conductivity8-32a}, the above-mentioned case corresponds to the point $d_k=0$ for the curves for $d_l=0$. Figure~\ref{fig:Conductivity8-32a}
clearly demonstrates that the electrical anisotropy  of the monolayer increases as the value of $k$ increases. 

\begin{figure}[htbp]
  \centering
  \includegraphics[width=\linewidth]{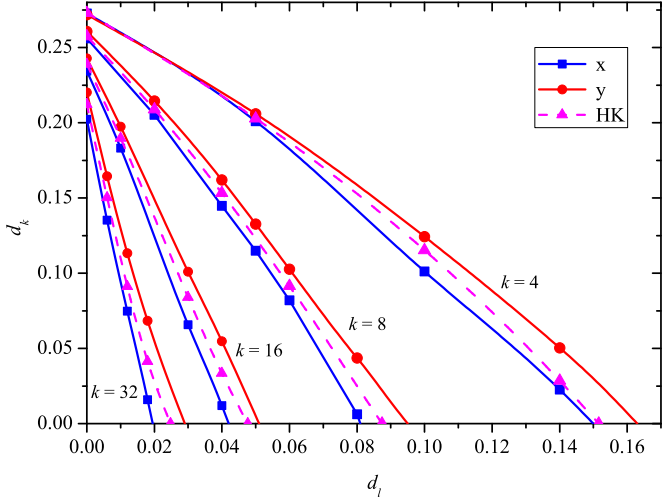}
  \caption{Phase diagram in the ($d_l,d_k$)-plane.
  Here, the solid lines for the $x$ and $y$ directions correspond to the effective  concentration at mean geometric conductivity $\sigma_g$, and the dashed lines were obtained using the Hoshen--Kopelman connectivity analysis at the thermodynamic limit~\cite{Hoshen1976PRB}. The anisotropic deposition, $k=4,8,16,32$, $L=100k$, results are averaged over 10 independent runs. The statistical error is of the order of the marker size. }\label{fig:PhaseDiagram}
\end{figure}

We found that defects in the lattice increase the  electrical anisotropy. At any given  value of $k$, the difference between the electrical conductivities along the $x$ and $y$ directions increases as the  concentration of defects $d_l$ grows. Insulating defects on the $k$-mers destroy connectivity when their concentration exceeds a  critical value. Figure~\ref{fig:PhaseDiagram} presents examples of phase diagrams in the  ($d_l,d_k$)-plane for $k=4,8,16,32$. The results for $k=2$ are omitted because the anisotropy of the electrical conductivity near the percolation threshold is negligible. Here, the solid lines correspond to the critical ``geometrical'' concentrations $d_k^x$ and $d_k^y$ for the  $x$ and $y$ directions, respectively, and the dashed lines were obtained using the Hoshen--Kopelman connectivity analysis at the thermodynamic limit~\cite{Hoshen1976PRB}. For each value of $k$, there is a conducting state when the concentration of defects is located below the curve. When  the concentrations is located above the curve, the monolayer is insulating. Quite surprising is the  region around the critical curve. This region corresponds to a monolayer with a high electrical conductivity along the $y$ direction and a low electrical conductivity along the $x$ direction.

\begin{figure}[htbp]
  \centering
\includegraphics[width=\linewidth]{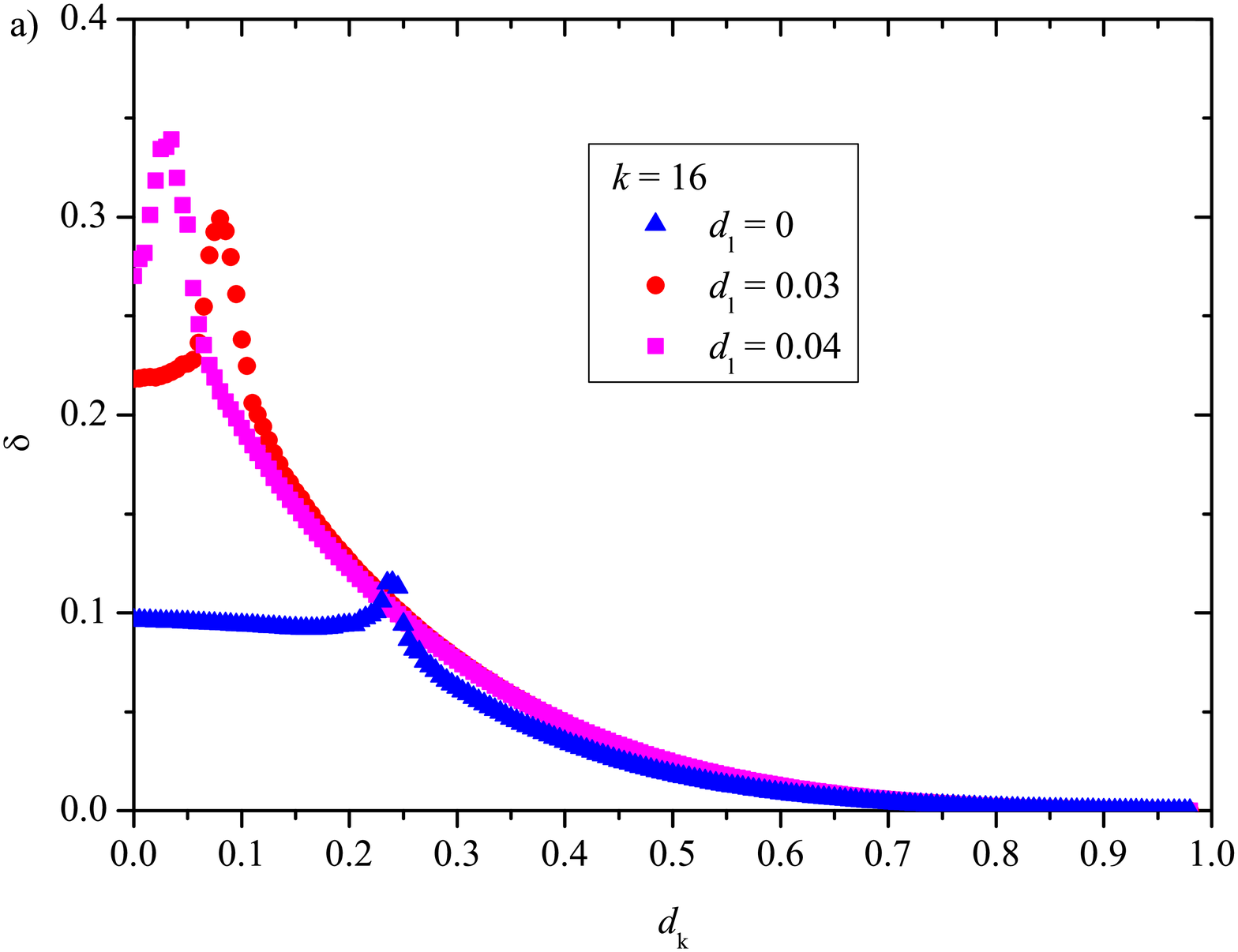}\\%
\includegraphics[width=\linewidth]{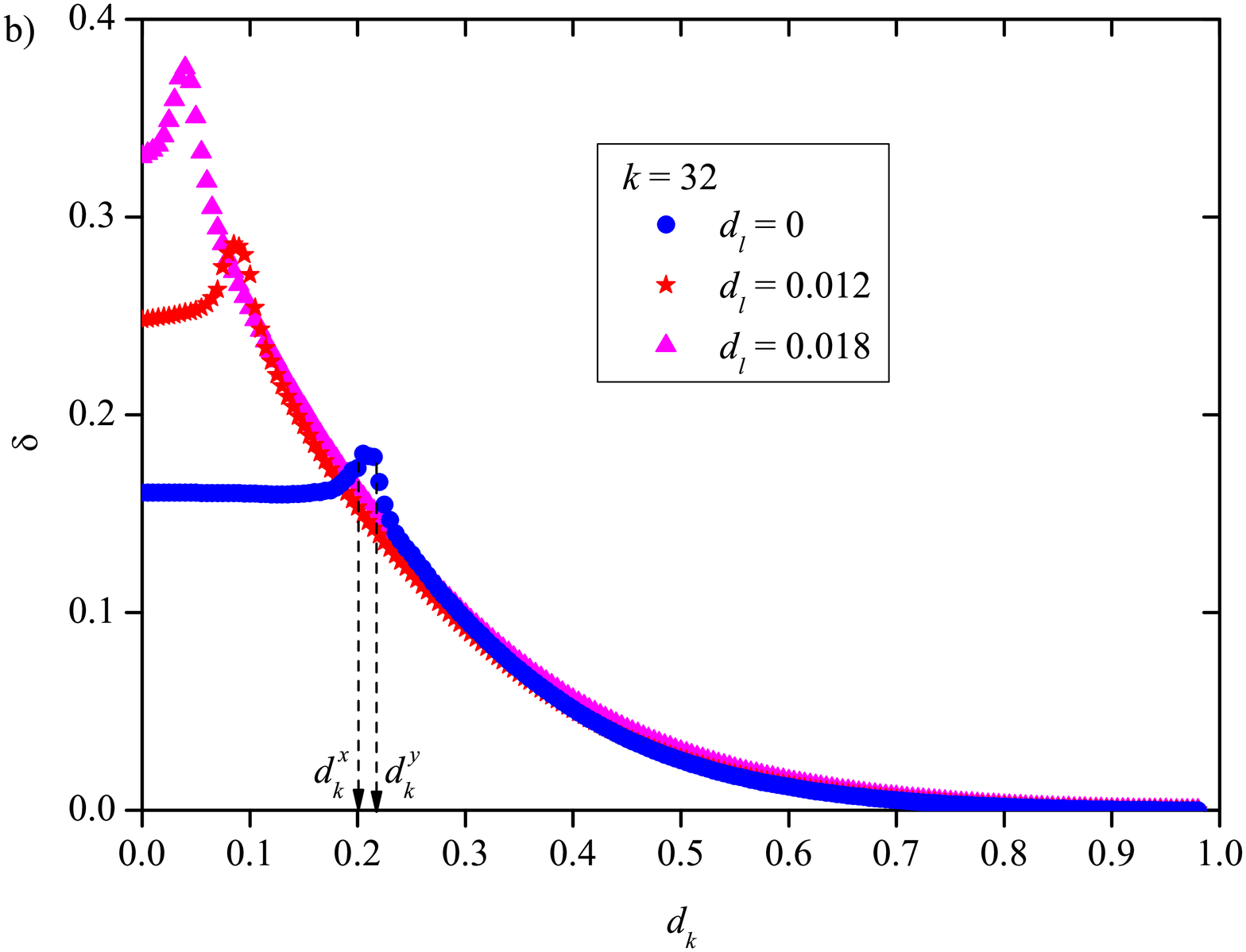}
  \caption{Anisotropy of the effective electrical conductivity, $\delta$, vs  defect concentration on the $k$-mers, $d_k$, for different values of defect concentration on a lattice, $d_l$. a) $k=16$, b) $k=32$. $L=100k$. Results are averaged over 10 independent runs, anisotropic deposition. The statistical error is of the  order of the marker size. The arrows  show the examples of the critical ``geometrical'' concentration $d_k^x$ and $d_k^y$ in the $x$ and $y$ directions.}\label{fig:Anisotropy}
\end{figure}

For characterization of the electrical anisotropy of monolayers, we used the same quantity as in~\cite{Tarasevich2016PRE}
\begin{equation}\label{eq:delta}
\delta =\frac{\lg \left(\sigma_x/\sigma_y\right)}{ \lg \left(\sigma_k/\sigma_l\right)}.
\end{equation}
This quantity equals 0 when a monolayer is electrically isotropic, and tends to 1 for a strongly  anisotropic  monolayer. For a defectless lattice ($d_l=0$), the anisotropy is large and constant for small values of $d_k$; it has a peak near the percolation threshold and tends to zero when the concentration of defects on the $k$-mers, $d_k$ increases (Figure~\ref{fig:Anisotropy}). The larger the value of $d_l$ the larger the initial anisotropy and the anisotropy near the percolation threshold, while the width of the initial plane part of the curve decreases. The effect is less pronounced for shorter particles (compare parts a) and b) in Figure~\ref{fig:Anisotropy}).

\section{Conclusion}\label{sec:conclusion}
In our research, the electrical conductivities of  monolayers of rod-like conducting particles adsorbed on an insulating substrate were calculated using the  Frank--Lobb algorithm~\cite{Frank1988PRB}. We considered both the anisotropic and  isotropic deposition of particles onto an imperfect substrate, i.e. a square lattice with embedded impurities. Calculation of the electrical conductivities gave an explicit confirmation of the predictions obtained on the basis of percolation theory~\cite{Lebovka2015PRE} viz. any kinds of defect have drastic negative effects on the electrical conductivity of the monolayers under consideration. In the case of anisotropic deposition, calculation showed that when the concentration of defects on the lattice, $d_l$, is insufficiently  large to destroy percolation, the monolayers with larger electrical anisotropy correspond to the larger values of $d_l$. Near the percolation threshold, the electrical anisotropy is greater. The evident anisotropy reflects the smearing of the percolation transition in the presence of an ``external field''~\cite{Snarskii2016}.
There are certain  concentrations of defects where the  sample is a good conductor along one direction and, at the same time, a bad conductor along the perpendicular direction. This observation suggests a means by which  samples exhibiting  high electrical anisotropy could be designed.

\acknowledgments

The reported study was supported by the Ministry of Education and Science of the Russian Federation, Project No.~643 and the National Academy of Sciences of Ukraine, Project No.~43/16-H.

\bibliography{conductivity,percolation}

\begin{thebibliography}{10}
\expandafter\ifx\csname url\endcsname\relax\def\url#1{\texttt{#1}}\fi

\bibitem{Maxwell1881}
\Name{Maxwell J.~C.} \Book{A treatise on electricity and magnetism} (Dover
  Publications) 1954.

\bibitem{Clerc1990AdPhys}
\Name{Clerc J.~P., Giraud G., Laugier J.~M. \and Luck J.~M.} \REVIEW{Adv.
  Phys.}{39}{1990}{191}.

\bibitem{Efros1976pssb}
\Name{Efros A.~L. \and Shklovskii B.~I.} \REVIEW{phys. status solidi
  (b)}{76}{1976}{475}.

\bibitem{McLachlan2007JNM}
\Name{McLachlan D.~S. \and Sauti G.} \REVIEW{J. Nanomater.}{2007}{2007}{1}.

\bibitem{Wang2008MSE}
\Name{Wang M. \and Pan N.} \REVIEW{Mater. Sci. Eng.: R: Reports}{63}{2008}{1}.

\bibitem{Snarskii2007PhysU}
\Name{Snarskii A.~A.} \REVIEW{Physics-Uspekhi}{50}{2007}{1239}.

\bibitem{Bruggeman1935AnnPhys}
\Name{Bruggeman D. A.~G.} \REVIEW{Ann. Phys.}{416}{1935}{636}.

\bibitem{Garboczi1991PRA}
\Name{Garboczi E.~J., Thorpe M.~F., DeVries M.~S. \and Day A.~R.} \REVIEW{Phys.
  Rev. A}{43}{1991}{6473}.

\bibitem{McLachlan1990JACE}
\Name{McLachlan D.~S., Blaszkiewicz M. \and Newnham R.~E.} \REVIEW{J. Am.
  Ceram. Soc.}{73}{1990}{2187}.

\bibitem{Balberg1983SSC}
\Name{Balberg I., Binenbaum N. \and Bozowski S.} \REVIEW{Solid State
  Commun.}{47}{1983}{989}.

\bibitem{Dykhne1971JETP}
\Name{Dykhne A.~M.} \REVIEW{Sov. Phys. --- JETP}{32}{1971}{63}.

\bibitem{Bulgadaev2003PLA}
\Name{Bulgadaev S.~A.} \REVIEW{Phys. Lett. A}{313}{2003}{106}.

\bibitem{Kondrat2006JChPh}
\Name{Kondrat G.} \REVIEW{J. Chem. Phys.}{124}{2006}{054713}.

\bibitem{BudinskiPetkovic2016JSM}
\Name{Budinski-Petkovi\'{c} L., Lon\v{c}arevi\'{c} I., Jak\v{s}i\'{c} Z.~M.
  \and Vrhovac S.~B.} \REVIEW{J. Stat. Mech. --- Theory
  E.}{2016}{2016}{053101}.

\bibitem{Centres2015JSMTE}
\Name{Centres P.~M. \and Ramirez-Pastor A.~J.} \REVIEW{J. Stat. Mech. ---
  Theory E.}{2015}{2015}{P10011}.

\bibitem{Hart2016PRE}
\Name{Hart R.~C. \and Aar\~ao Reis F. D.~A.} \REVIEW{Phys. Rev.
  E}{94}{2016}{022802}.

\bibitem{Kuriata2016MTS}
\Name{Kuriata A., Polanowski P. \and Sikorski A.} \REVIEW{Macromol. Theor.
  Simul.}{25}{2016}{360}.

\bibitem{Wisniowski2016PhysA}
\Name{Wi\'{s}niowski R., Olchawa W., Fr\c{a}czek D. \and Piasecki R.}
  \REVIEW{Physica A}{444}{2016}{799}.

\bibitem{Cherkasova2010EPJB}
\Name{Cherkasova V.~A., Tarasevich Y.~Y., Lebovka N.~I. \and Vygornitskii
  N.~V.} \REVIEW{Eur. Phys. J. B}{74}{2010}{205}.

\bibitem{Tarasevich2016PRE}
\Name{Tarasevich Y.~Y., Goltseva V.~A., Laptev V.~V. \and Lebovka N.~I.}
  \REVIEW{Phys. Rev. E}{94}{2016}{042112}.

\bibitem{Adamson1997}
\Name{Adamson A. \and Gast A.} \Book{Physical chemistry of surfaces} (Wiley)
  1997.

\bibitem{Cadilhe2007JPhysCM}
\Name{Cadilhe A., Ara\'{u}jo N. A.~M. \and Privman V.} \REVIEW{J. Phys. Cond.
  Matt.}{19}{2007}{065124}.

\bibitem{Ritter2010}
\Name{Ritter U., Scharff P., Pinchuk T., Dmytrenko O., Bulavin L., Kulish M.,
  Prylutskyy Y.~I., Zabolotnyy M., Grabovsky Y.~E., Bilyy M., Rugal A., Shut A.
  \and Shlapatska V.} \REVIEW{Materialwissenschaft und
  Werkstofftechnik}{41}{2010}{675}.

\bibitem{Wepasnick2010}
\Name{Wepasnick K.~A., Smith B.~A., Bitter J.~L. \and Fairbrother D.~H.}
  \REVIEW{Anal. Bioanal. Chem.}{396}{2010}{1003}.

\bibitem{Ben-Naim1994JPhysA}
\Name{Ben-Naim E. \and Krapivsky P.~L.} \REVIEW{J. Phys. A}{27}{1994}{3575}.

\bibitem{Lee1996JPhysA}
\Name{Lee J.~W.} \REVIEW{J. Phys. A}{29}{1996}{33}.

\bibitem{Kondrat2005}
\Name{Kondrat G.} \REVIEW{J. Chem. Phys.}{122}{2005}{184718}.

\bibitem{Cornette2003epjb}
\Name{Cornette V., Ramirez-Pastor A. \and Nieto F.} \REVIEW{Eur. Phys. J.
  B}{36}{2003}{391}.

\bibitem{Cornette2006PLA}
\Name{Cornette V., Ramirez-Pastor A. \and Nieto F.} \REVIEW{Phys. Lett.
  A}{353}{2006}{452}.

\bibitem{Cornette2011PhysA}
\Name{Cornette V., Ramirez-Pastor A. \and Nieto F.} \REVIEW{Phys.
  A}{390}{2011}{671}.

\bibitem{Budinski-Petkovic2011}
\Name{Budinski-Petkovi\'{c} L., Lon\v{c}arevi\'{c} I., Jak\v{s}i\'{c} Z.~M.,
  Vrhovac S.~B. \and \v{S}vraki\'{c} N.~M.} \REVIEW{Phys. Rev.
  E}{84}{2011}{051601}.

\bibitem{Budinski-Petkovic2012}
\Name{Budinski-Petkovi\'{c} L., Lon\v{c}arevi\'{c} I., Petkovi\'{c} M.,
  Jak\v{s}i\'{c} Z.~M. \and Vrhovac S.~B.} \REVIEW{Phys. Rev.
  E}{85}{2012}{061117}.

\bibitem{Tarasevich2015PRE}
\Name{Tarasevich Y.~Y., Laptev V.~V., Vygornitskii N.~V. \and Lebovka N.~I.}
  \REVIEW{Phys. Rev. E}{91}{2015}{012109}.

\bibitem{Lebovka2015PRE}
\Name{Lebovka N.~I., Tarasevich Y.~Y., Dubinin D.~O., Laptev V.~V. \and
  Vygornitskii N.~V.} \REVIEW{Phys. Rev. E}{92}{2015}{062116}.

\bibitem{Tarasevich2016JPhCS}
\Name{Tarasevich Y.~Y., Dubinin D.~O., Laptev V.~V. \and Lebovka N.~I.}
  \REVIEW{J. Phys.: Conf. Ser.}{681}{2016}{012038} {International Conference on
  Computer Simulation in Physics and Beyond 2015}.

\bibitem{Evans1993RMP}
\Name{Evans J.~W.} \REVIEW{Rev. Mod. Phys.}{65}{1993}{1281}.

\bibitem{Frank1988PRB}
\Name{Frank D.~J. \and Lobb C.~J.} \REVIEW{Phys. Rev. B}{37}{1988}{302}.

\bibitem{Snarskii2016}
\Name{Snarskii A.~A., Bezsudnov I.~V., Sevryukov V.~A., Morozovskiy A. \and
  Malinsky J.} \Book{Transport Processes in Macroscopically Disordered Media:
  From Mean Field Theory to Percolation} (Springer New York, New York, NY) 2016
  Ch. Effective Conductivity of Percolation Media pp. 47--75.

\bibitem{Hoshen1976PRB}
\Name{Hoshen J. \and Kopelman R.} \REVIEW{Phys. Rev. B}{14}{1976}{3438}.

\end{thebibliography}

\end{document}